\documentclass[galaxies,article,accept,pdftex,moreauthors]{Definitions/mdpi} 

\firstpage{1} 
\pubvolume{1}
\issuenum{1}
\articlenumber{0}
\pubyear{2024}
\copyrightyear{2024}
\datereceived{ } 
\daterevised{ } 
\dateaccepted{ } 
\datepublished{ } 
\hreflink{https://doi.org/} 

\Title{A 3~mm Spectral Line Study of the Central Molecular Zone Infrared Dark Cloud G1.75-0.08}

\TitleCitation{A 3~mm Spectral Line Study of the Central Molecular Zone Infrared Dark Cloud G1.75-0.08}

\Author{Oskari 
 Miettinen $^{1,*}$\orcidA{} and Miguel Santander-Garc\'{\i}a $^{2,3}$\orcidB{}}

\AuthorNames{Oskari Miettinen and Miguel Santander-Garc\'{\i}a}

\AuthorCitation{Miettinen, 
 O.; Santander-\linebreak Garc\'{\i}a, M.}

\address{%
$^{1}$ \quad Research Council of Finland, Hakaniemenranta 6, P.O. Box 131, FI-00531 Helsinki, Finland\\
$^{2}$ \quad Observatorio Astron\'omico Nacional (OAN-IGN), Alfonso XII, 3, 28014 Madrid, Spain; m.santander@oan.es\\
$^{3}$ \quad Centro de Desarrollos Tecnol\'ogicos, Observatorio de Yebes (IGN), 19141 Yebes, Guadalajara, 
 Spain}

\corres{Correspondence: oskari.miettinen@aka.fi}

\abstract{Infrared dark clouds (IRDCs) are fruitful objects to study the fragmentation of interstellar filaments and initial conditions and early stages of high-mass ($M>8$~M$_{\odot}$) star formation. We used the Yebes 40~m and Institut de Radioastronomie Millim\'etrique (IRAM) 30~m radio telescopes to carry out the first single-pointing spectral line observations towards the IRDC G1.75-0.08, which is a filamentary Central Molecular Zone (CMZ) cloud. Our aim is to reach an improved understanding of the gas kinematics and dynamical state of the cloud and its two clumps that we call clumps~A and B. We also aim to determine the fractional abundances of the molecules detected at 3~mm towards G1.75-0.08. We detected HNCO$(J_{K_a,\,K_c}=4_{0,\,4}-3_{0,\,3})$, HCN$(J=1-0)$, and HCO$^+(J=1-0)$ towards both clumps. The N$_2$H$^+(J=1-0)$ line was detected only in clump~B, while N$_2$D$^+(J=1-0)$ was not detected at all. The HCN and HNCO spectra exhibit two velocity components. The abundances of the detected species are comparable to those in other IRDCs. An upper limit to the [N$_2$D$^+$]/[N$_2$H$^+$] deuterium fraction of $<0.05$ derived towards clump~B is consistent with values observed in many high-mass clumps. The line mass analysis suggests that the G1.75-0.08 filament is subcritical by a factor of $11\pm6$, and the clumps were found to be gravitationally unbound ($\alpha_{\rm vir} > 2$). Our finding that G1.75-0.08 is strongly subcritical is atypical compared to the general population of Galactic filamentary clouds. The cloud's location in the CMZ might affect the cloud kinematics similar to what has been found for the Brick IRDC, and the cloud's dynamical state might also be the result of the turbulent motions or shear and tidal forces in the CMZ. Because the target clumps are dark at 70~$\upmu$m and massive (several $10^3$~M$_{\odot}$), they can be considered candidates for being high-mass starless clumps but not prestellar because they are not gravitationally bound.}


\begin{document}




\section{Introduction}

Interstellar infrared dark clouds (IRDCs) are identified as dark absorption features against the Galactic mid-IR background radiation 
(e.g.,~\cite{perault1996, egan1998, peretto2009}). Infrared dark clouds are very useful target sources for the studies of molecular cloud fragmentation and star formation. One reason for this is that IRDCs often exhibit filamentary morphology with substructure along the long axis of the filament (e.g.,~\cite{jackson2010, kainulainen2013, henshaw2016, miettinen2018}). Another reason for IRDCs being useful target sources is that some of them show evidence of ongoing high-mass ($M>8$~M$_{\odot}$) star formation (e.g.,~\cite{rathborne2006, beuther2007, chambers2009, battersby2010, retes2020}), a~process where many details and even the exact mechanism(s) are still to be deciphered (e.g.,~\cite{motte2018} for a review). On~the other hand, IRDCs are promising sites to search for high-mass prestellar clumps and cores (e.g.,~\cite{sanhueza2017}).

\begin{figure}[H]
\includegraphics[width=0.6\textwidth]{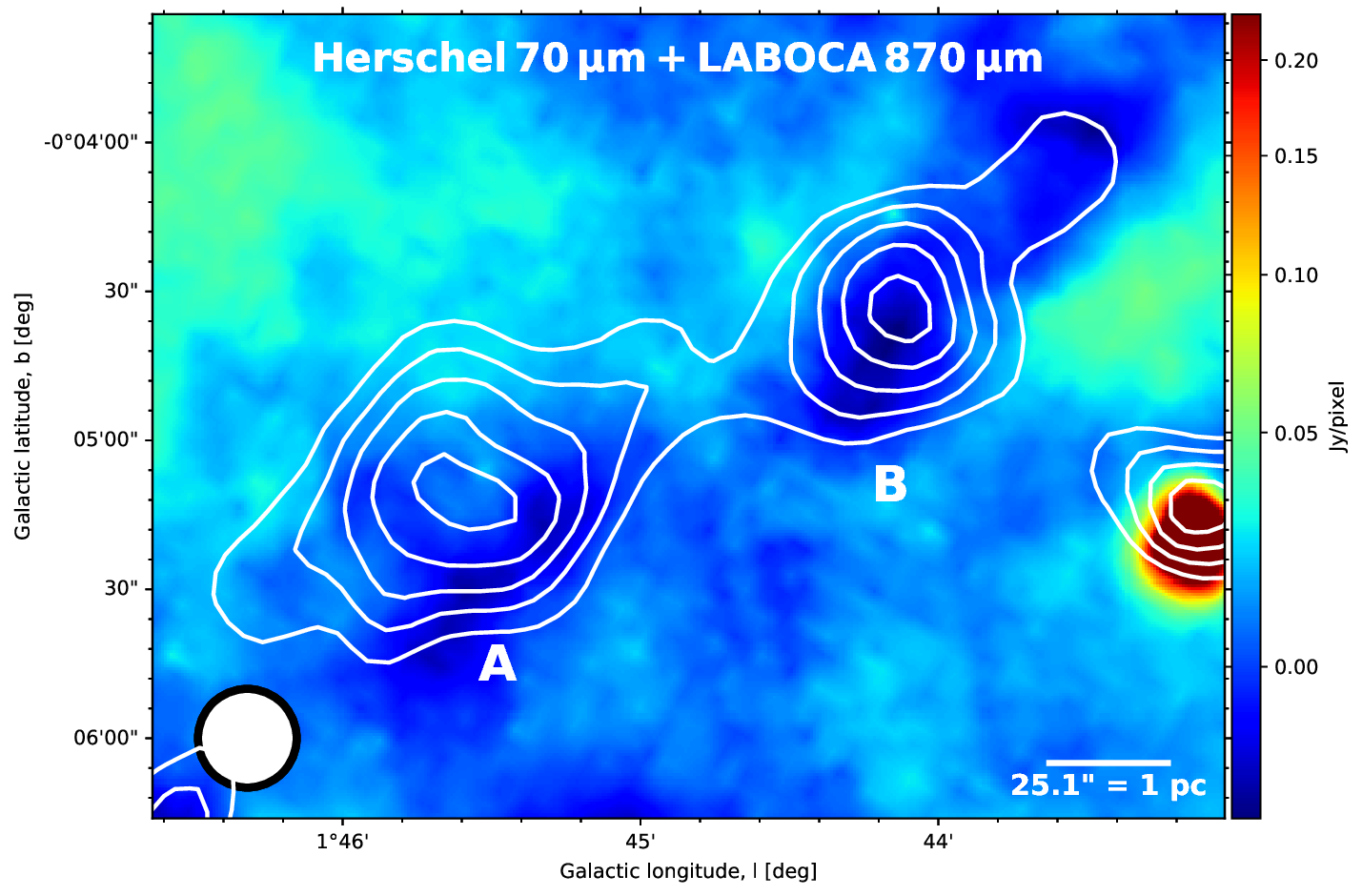}
\caption{\textit{Herschel 
} 70~$\upmu$m image towards G1.75-0.08 (colour map) overlaid with contours showing the LABOCA 870~$\upmu$m emission. The~contour levels start from $3\sigma$ and progress in steps of $1\sigma$, where $1\sigma=90$~mJy~beam$^{-1}$. The~two clumps, clump~A and clump~B, are labelled A and B. The~beam size (HPBW of 19.9 arcsec) of the LABOCA observations is shown in the bottom left corner and a scale bar of 1~pc is shown in the bottom right~corner.\label{figure:map}}
\end{figure}

In this paper, we present a molecular spectral line study of the Galactic IRDC G1.75-0.08 (Figure~\ref{figure:map}). The~target IRDC was uncovered by dust continuum imaging survey with the Large APEX BOlometer CAmera (LABOCA;~\cite{siringo2009}) at 870~$\upmu$m by Miettinen~\cite{miettinen2012}, and~it was part of the sample in the molecular line study of IRDCs by Miettinen~\cite{miettinen2014} that was based on the Millimetre Astronomy Legacy Team 90~GHz (MALT90) survey~\cite{foster2011, foster2013, jackson2013}. A~follow-up dust continuum imaging study of G1.75-0.08 at 350~$\upmu$m and 450~$\upmu$m was conducted by Miettinen~et~al.~\cite{miettinen2022} using the Architectures de bolom\`{e}tres pour des T\'elescopes \`{a} grand champ de vue dans le domaine sub-Millim\'etrique au Sol, or~ArT\'eMiS bolometer~\cite{reveret2014, andre2016, talvard2018}. Miettinen~et~al.~\cite{miettinen2022} revised the kinematic distance of G1.75-0.08 to be 8.22~kpc with a Galactocentric distance of 270~pc. Hence, G1.75-0.08 is located within the Central Molecular Zone (CMZ) of the Galaxy, where the cloud can be associated with extreme environmental conditions such as strong tidal shear forces, strong turbulence, and~more complex chemistry than found in the Galactic spiral arm molecular clouds (e.g.,~\cite{petkova2023}; see also~\cite{henshaw2023} for a review).

Prior to the present study, the~only spectral line data available for G1.75-0.08 were those obtained as part of the MALT90 survey. The~modest sensitivity of the MALT90 data ($\sim$250~mK per 0.11~km~s$^{-1}$ channel) and the angular resolution (half-power beam width, or~HPBW) of those data, 38 arcsec, complicated the interpretation of the few extracted line detections (e.g., HCN$(J=1-0)$ and HCO$^+(J=1-0)$; \cite{miettinen2014}). Our new spectral line observations with the Yebes 40~m telescope benefit from 1.9 times higher angular resolution compared to the MALT90 data, which is useful owing to the large distance of the cloud from the Sun. In~the present paper, we will revise some of the source properties that depend on information provided by spectral line data (in particular, spectral line widths), and~determine the fractional abundances of the molecules detected towards the target source via our 3~mm band~observations.

The observations and data reduction are described in Section~\ref{sec2}. The~analysis and results are described in Section~\ref{sec3}. We discuss the results 
in Section~\ref{sec4}, while in Section~\ref{sec5}, we summarise our results and main~conclusions. 

\unskip

\section{Observations and Data~Reduction}\label{sec2}
\unskip

\subsection{Yebes~Observations}\label{sec2.1}

We used the Yebes 40~m radio telescope~\cite{tercero2021} to observe the LABOCA 870~$\upmu$m peak positions of the clumps in G1.75-0.08 at a frequency range of about 72--90.5~GHz. The observed frequency was tuned at 80.05~GHz. 
 The target clumps are listed in Table~\ref{table:G175sources}. 

Our observations made use of the W band, where the full bandwidth is 18.5~GHz. The~spectral backend was a Fast-Fourier Transform Spectrometer (FFTS) with a spectral resolution of 38~kHz, and~the observations were made in dual (horizontal and vertical) linear polarisation mode. The~aforementioned spectral resolution corresponds to a velocity resolution of 126--158.3~m~s$^{-1}$ over the observed frequency range. The~beam size (HPBW) at the observed frequency range is 19.5--24.5 arcsec (\cite{tercero2021}, Table~5 therein).   

\begin{table}[H] 
\caption{Target 
 clumps in~G1.75-0.08.\label{table:G175sources}}
\newcolumntype{C}{>{\centering\arraybackslash}X}
\begin{tabularx}{\textwidth}{CCCCC}
\toprule
\textbf{Source} & \boldmath{$\alpha_{2000.0}$} & \boldmath{$\delta_{2000.0}$} & \boldmath{$T_{\rm dust}$} & \boldmath{$M$}\\
       & \textbf{[h:m:s]} & \textbf{[}\boldmath{$^{\circ}$:$^{'}$:$^{''}$}\textbf{]} &\textbf{ [K]} & \textbf{[M}\boldmath{$_{\odot}$}\textbf{]} \\
\midrule
Clump A & 17 50 05.08 & $-$27 28 32.12 & $14.3\pm0.4$ & $5\,501\pm935$ \\
Clump B & 17 49 59.38 & $-$27 29 24.98 & $14.5\pm0.5$ & $3\,770\pm738$ \\
\bottomrule
\end{tabularx}
\noindent{\footnotesize{The columns are as follows: (1) source name; (2) and (3) coordinates of the LABOCA 870~$\upmu$m peak surface brightness position; (4) dust temperature; (5) total (gas+dust) mass of the clump. The~values of $T_{\rm dust}$ and $M$ were derived by Miettinen~et~al.~\cite{miettinen2022}.}}
\end{table}

\vspace{-4pt}
The first observations were carried out on 27 and 28 March 2023 (for clump~B and clump~A, respectively) under the project 23A004 
(PI: Miettinen). These observations were performed in the position switching mode, where the off position for clump~A was 1.092~arcmin to the north of the on position, and~for clump~B, the reference position was located at 0.874 arcmins to the north of the target position. These offsets were chosen on the basis of the extent of the LABOCA 870~$\upmu$m emission (Figure~\ref{figure:map}). The~total on-source integration time for clump~A was 31~min, and for clump~B, it was about 21~min. However, it was found that the reference off positions of the observations had spectral line emission at the observed frequencies, which resulted in absorption-like features in the final spectra that corrupted the lines observed towards the on~positions.

The observations were repeated on 11 and 16 February 2024 (for clump~A and clump~B, respectively) under project 24A005 (PI: Miettinen). These observations were performed in the frequency switching mode with a conservative frequency throw of 74.1~MHz ($\pm37.05$~MHz frequency offsets). 
Because, in the frequency switching mode, the observed position does not change, we could avoid the possible presence of spectral line emission in the position switching mode's reference position. The~pointing and focus corrections were conducted by observing a SiO maser line at 86,243.4277~MHz towards the Red Supergiant VX Sgr. The~total on-source integration time for clump~A was 1.15~h, and for clump~B, it was 1.9~h.

The system temperature during the observations towards clump~A was $T_{\rm sys}\simeq200$~K, while for observations of clump~B, it was about 300~K. Calibration was made by the hot-cold load technique, and~the output intensity scale given by the system is the antenna temperature corrected for the atmospheric attenuation ($T_{\rm A}^*$). The~observed intensities were converted to the main-beam brightness temperature scale using the formula $T_{\rm MB}=T_{\rm A}^*/\eta_{\rm MB}$, where $\eta_{\rm MB}$ is the main-beam efficiency. The~values of $\eta_{\rm MB}$ range from 0.20 to 0.31 over the observed frequency range.\endnote{\url{https://rt40m.oan.es/rt40m_en.php} accessed on 1 February 2024
.} The absolute calibration uncertainty was adopted to be 10\% following the practice of studies that employ the Yebes Q-band observations (e.g.,~\cite{silva2023, agundez2023, tercero2024}). However, this should be taken as a lower limit to the calibration uncertainty in the higher frequency W band. Moreover, it has been observed that the spectral line intensity discrepancies between position and frequency switching observations with the Yebes 40~m radio telescope are always $<20\%$ \cite{tercero2021}. 

The spectra were reduced using Continuum and Line Analysis Single-dish Software 
 ({\tt CLASS90}) of the GILDAS software package (version jul21)\endnote{Grenoble Image and Line Data Analysis Software (GILDAS) is provided and actively developed by IRAM, and~is available at \url{www.iram.fr/IRAMFR/GILDAS} accessed on 28 December 2022
.}. The~individual spectra were stitched with weights defined as $w_{\rm i} \propto t_{\rm int}/T_{\rm sys}^2$, where $t_{\rm int}$ is the integration time. The~resulting spectra were folded and then smoothed using the Hann window function so that the number of channels was divided by two. First or second-order baselines were determined from the velocity ranges free of spectral line features and then subtracted from the spectra. The~resulting $1\sigma$ rms noise levels at the smoothed velocity resolution were about 20--33~mK with an average of 26.6~mK on a $T_{\rm A}^*$ scale.

\subsection{IRAM~Observations}

We also used the Institut de Radioastronomie Millim\'etrique (IRAM) 30 m telescope to observe the target positions (i.e., the LABOCA 870~$\upmu$m peaks of the clumps) in the $J=1-0$ transition of N$_2$H$^+$ and N$_2$D$^+$. The~observations were carried out on 8 and 14 February 2024 during a pool observing week (project 079-23; PI: Miettinen).

The Eight MIxer Receiver (EMIR;~\cite{carter2012}) band E090 was used as a front end, while the backend was the Versatile SPectrometer Array (VESPA), where the bandwidth was $18 \times 20$~MHz with a corresponding channel spacing of 20~kHz. The~N$_2$H$^+(1-0)$ and N$_2$D$^+(1-0)$ lines were 
tuned at the frequencies of the strongest hyperfine component ($J_{F_1,\, F} = 1_{2,\, 3}-0_{1,\, 2}$) of the transition, which are $93,173.7637$~MHz 
and 77,109.6162~MHz, respectively, (\cite{pagani2009}, Tables~2 and 8 therein). 

The aforementioned channel spacing yielded a velocity resolution of about 64~m~s$^{-1}$ for N$_2$H$^+$ and 78~m~s$^{-1}$ for N$_2$D$^+$. The~telescope beam sizes (HPBW) at the observed frequencies are 26.4 arcsec (N$_2$H$^+$) and 31.9 arcsec (N$_2$D$^+$).

The observations were performed in the frequency switching
mode with a frequency throw of $\pm3.9$~MHz. The~N$_2$H$^+(1-0)$ observing time for 
both clumps was about 1.1~h. The~N$_2$D$^+(1-0)$ observing time for clump~A was 2.2~h and 1.6~h for clump~B.
The telescope focus and pointing were optimised and checked on the quasars 1226+023 (3C 273) and 1757-240.
The precipitable water vapour (PWV) during the observations was measured to be between 4.2 mm and 5.5 mm. The~system temperatures during
the observations were within the range of $T_{\rm sys}=113-140$~K.
The observed intensities were converted into the main-beam
brightness temperature scale by using a main-beam efficiency factor of $\eta_{\rm MB}=0.80$ for N$_2$H$^+(1-0)$ and 0.81 for N$_2$D$^+(1-0)$. The~absolute calibration uncertainty was adopted to be 10\% (e.g.,~\cite{desimone2018}).

The spectra were reduced using {\tt CLASS90} (version jul21) in a similar fashion as described in Section~\ref{sec2.1} (including smoothing that halves the spectral resolution). Second- or third-order polynomial baselines were determined from the velocity ranges free of spectral line features, and~then subtracted from the spectra. The~resulting $1\sigma$ rms noise levels were about 13–21~mK.

\section{Analysis and~Results}\label{sec3}
\unskip

\subsection{Detected Spectral Lines and Spectral Line~Parameters}\label{sec3.1}

\begin{figure}[H]
\begin{center}
\includegraphics[width=0.32\textwidth]{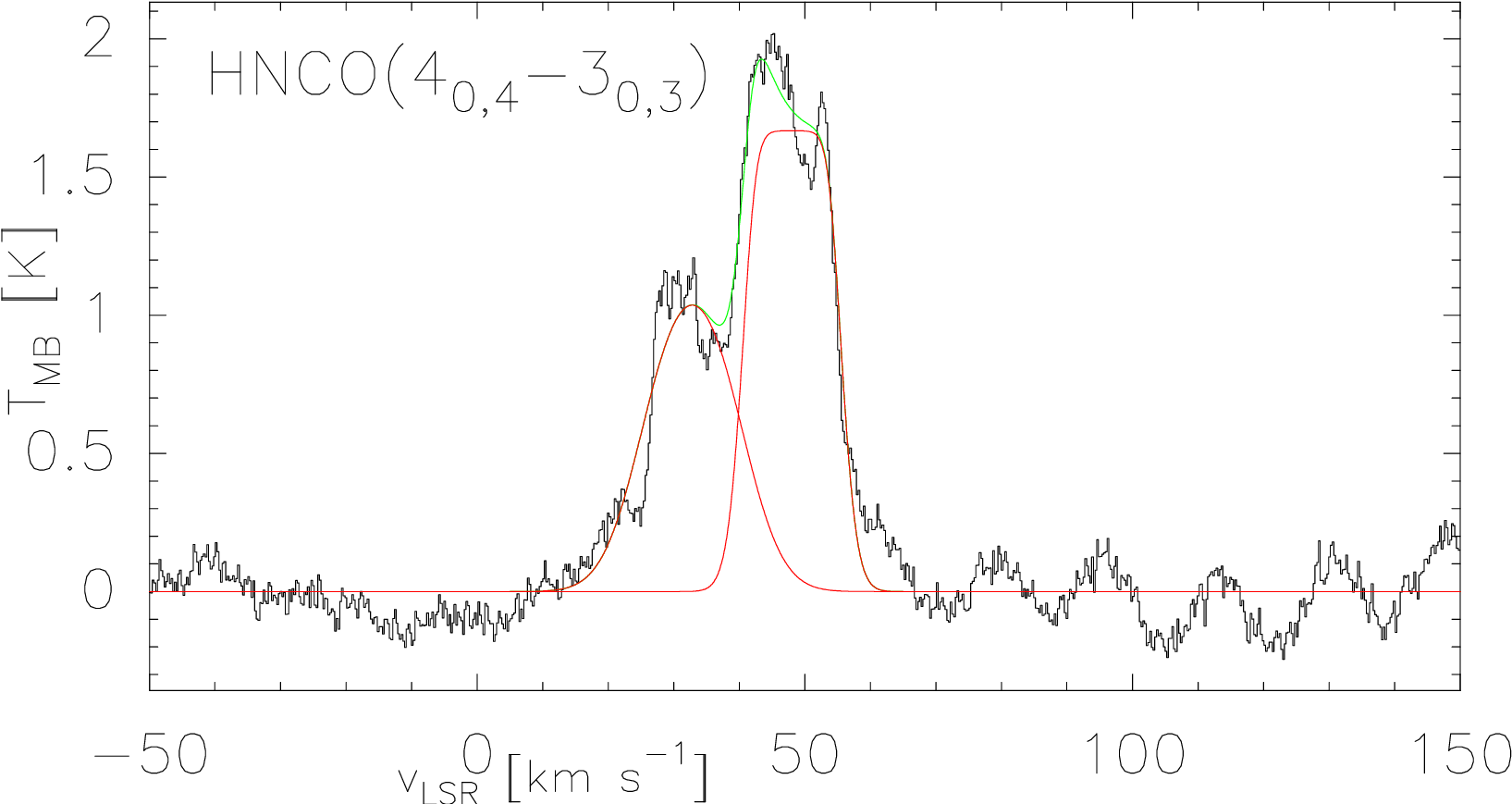}
\includegraphics[width=0.32\textwidth]{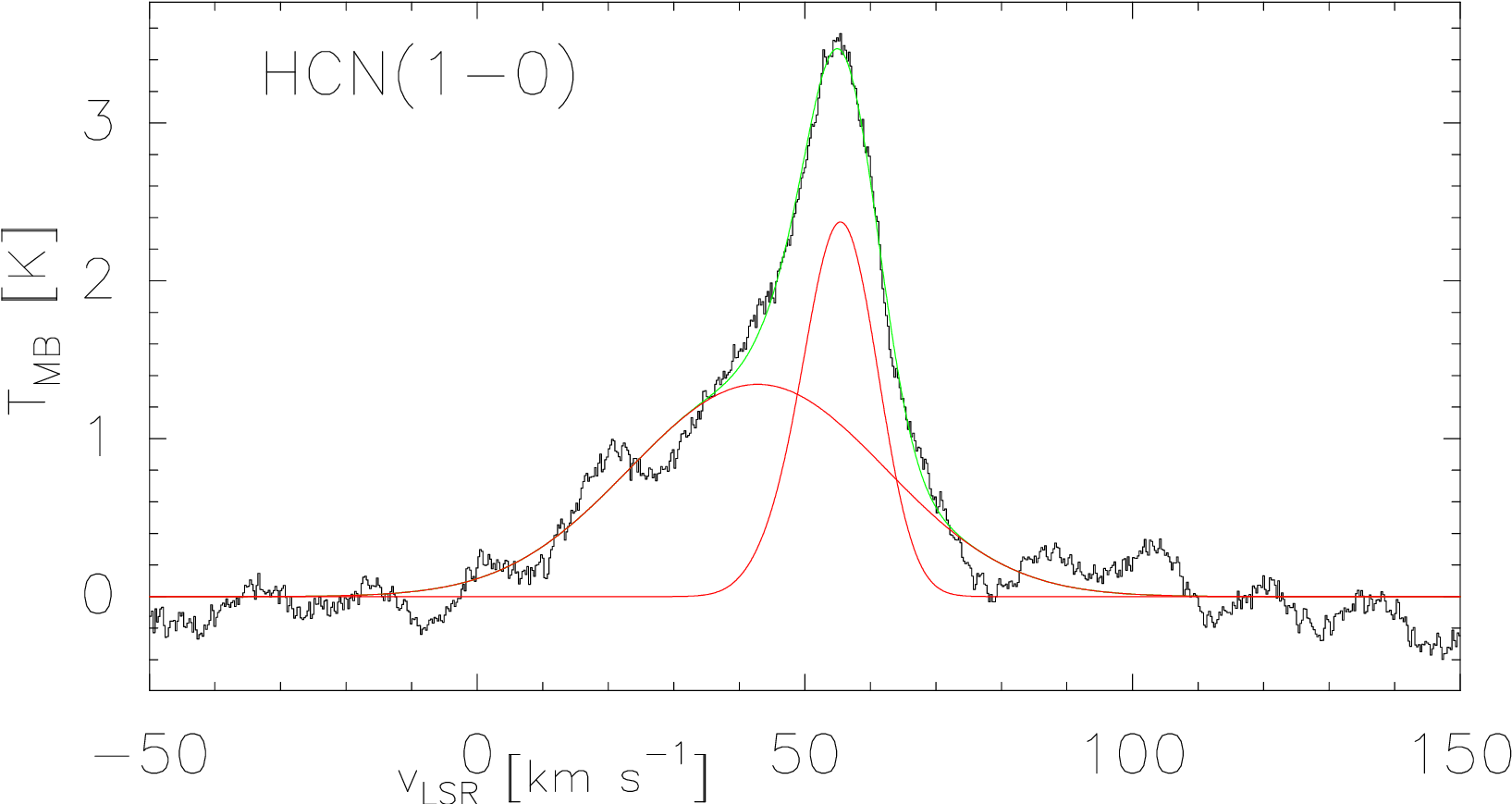}
\includegraphics[width=0.32\textwidth]{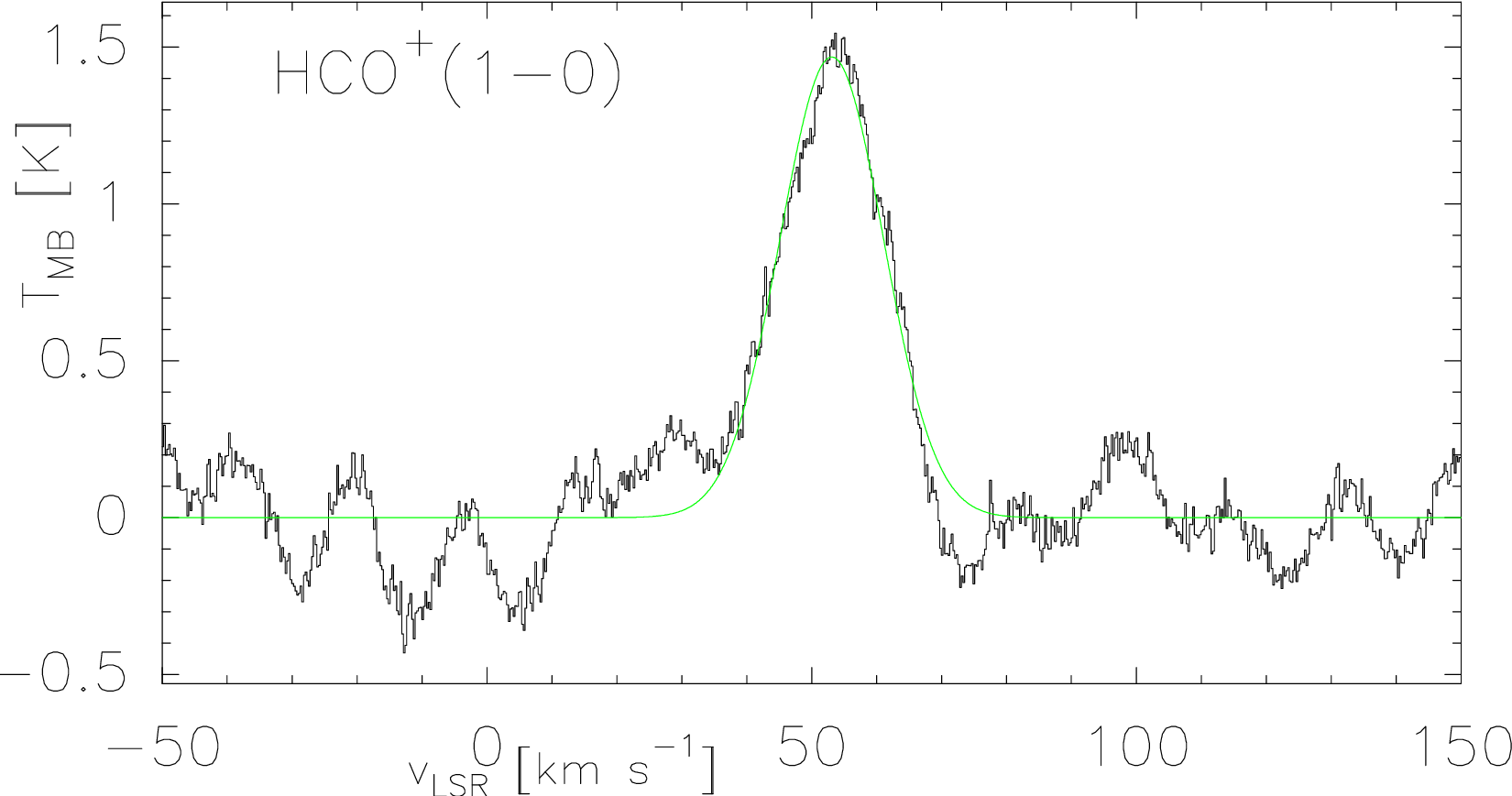}
\caption{Spectral 
 lines detected with the Yebes telescope towards clump~A. Hyperfine structure fits to the two velocity components in the HNCO and HCN spectra are shown in red, while the green line shows the sum of the two components. A~single Gaussian fit to the HCO$^+$ line is overlaid in green. The~intensity range in the panels differs from each other to better show the line~profiles.\label{figure:spectra1}}
\end{center}
\end{figure}

\begin{figure}[H]
\includegraphics[width=0.32\textwidth]{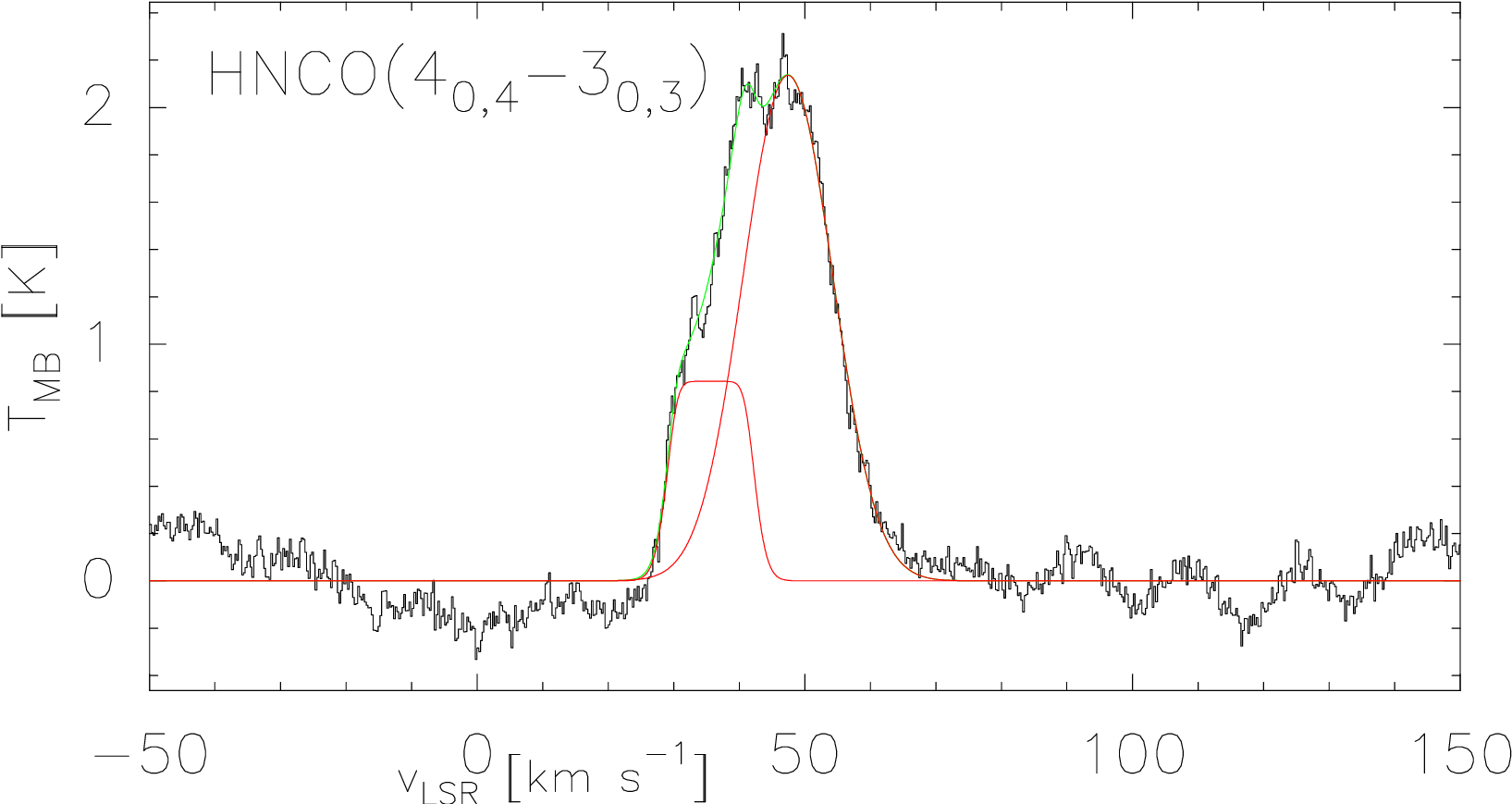}
\includegraphics[width=0.32\textwidth]{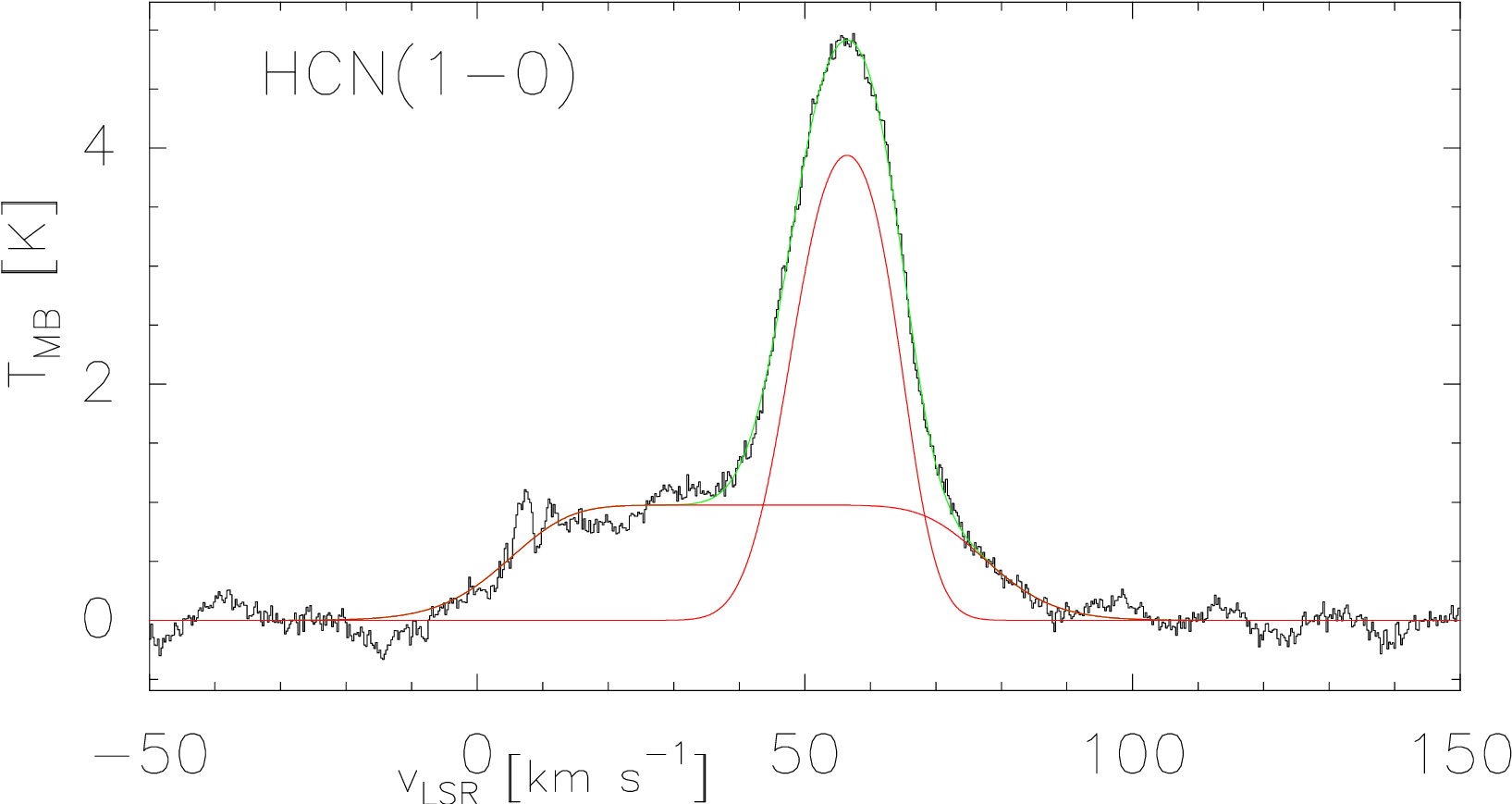}
\includegraphics[width=0.32\textwidth]{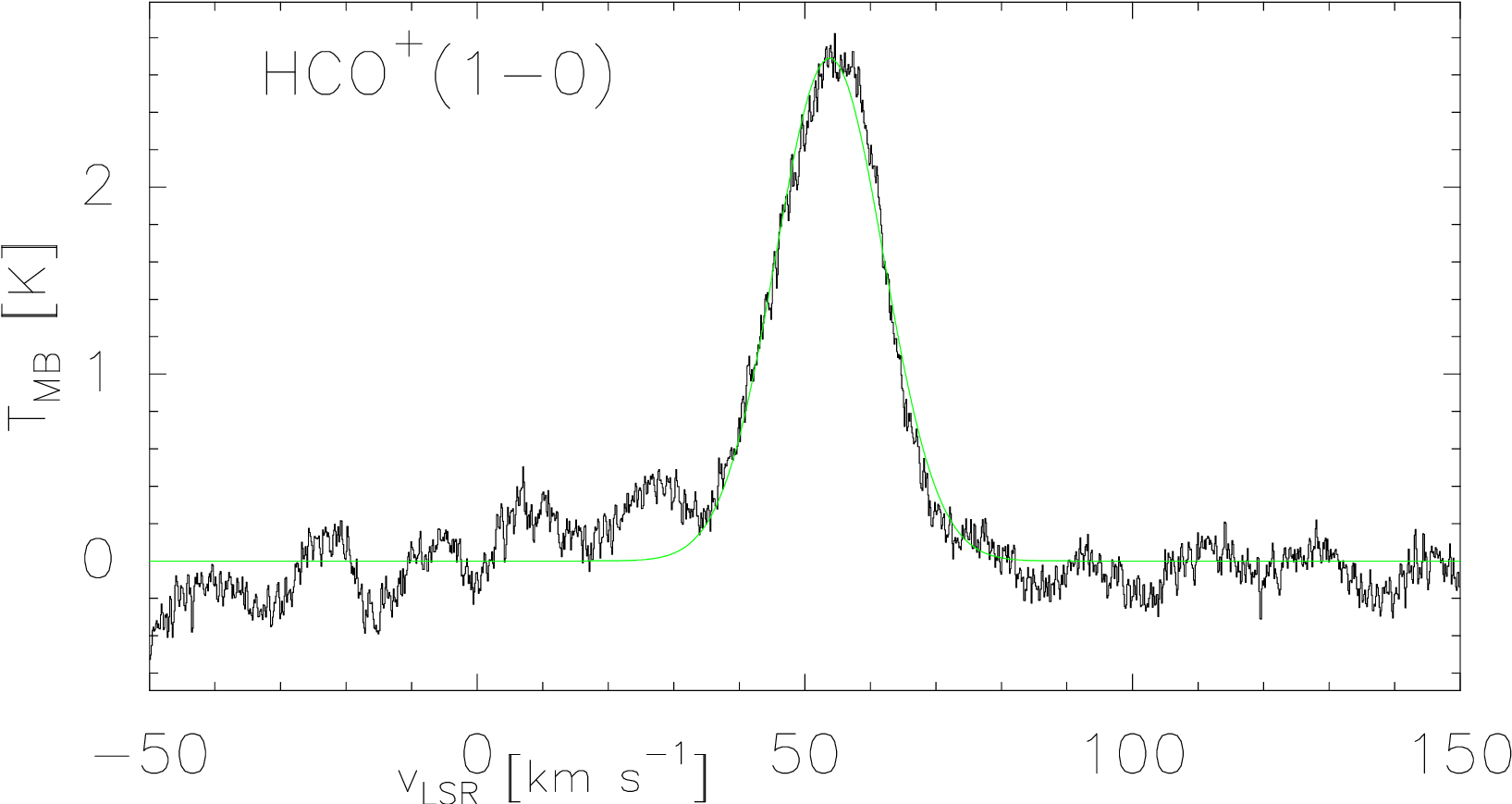}
\caption{Same as Figure~\ref{figure:spectra1} but for clump~B.\label{figure:spectra2}}
\end{figure}

\begin{figure}[H]
\includegraphics[width=0.41\textwidth]{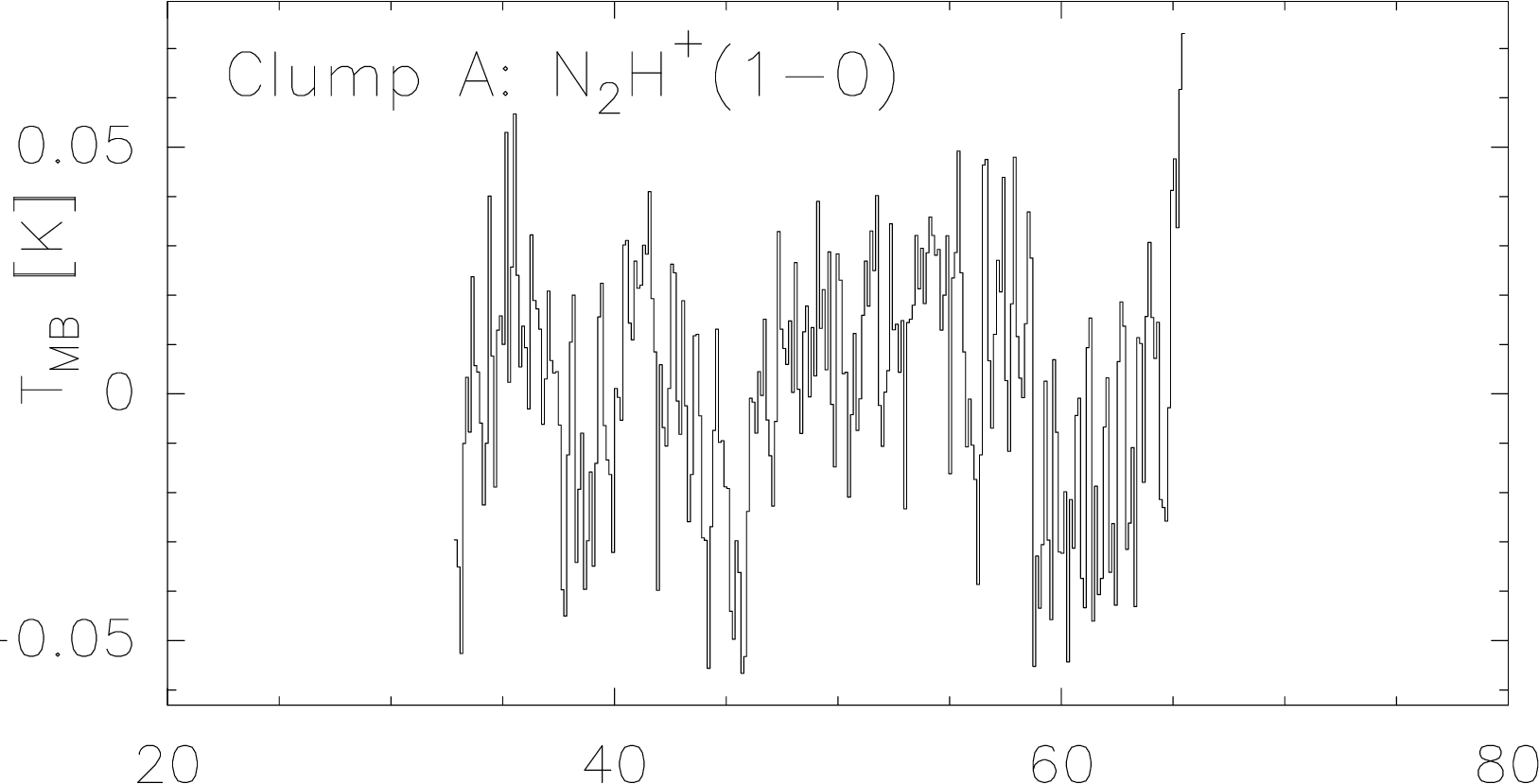}
\includegraphics[width=0.41\textwidth]{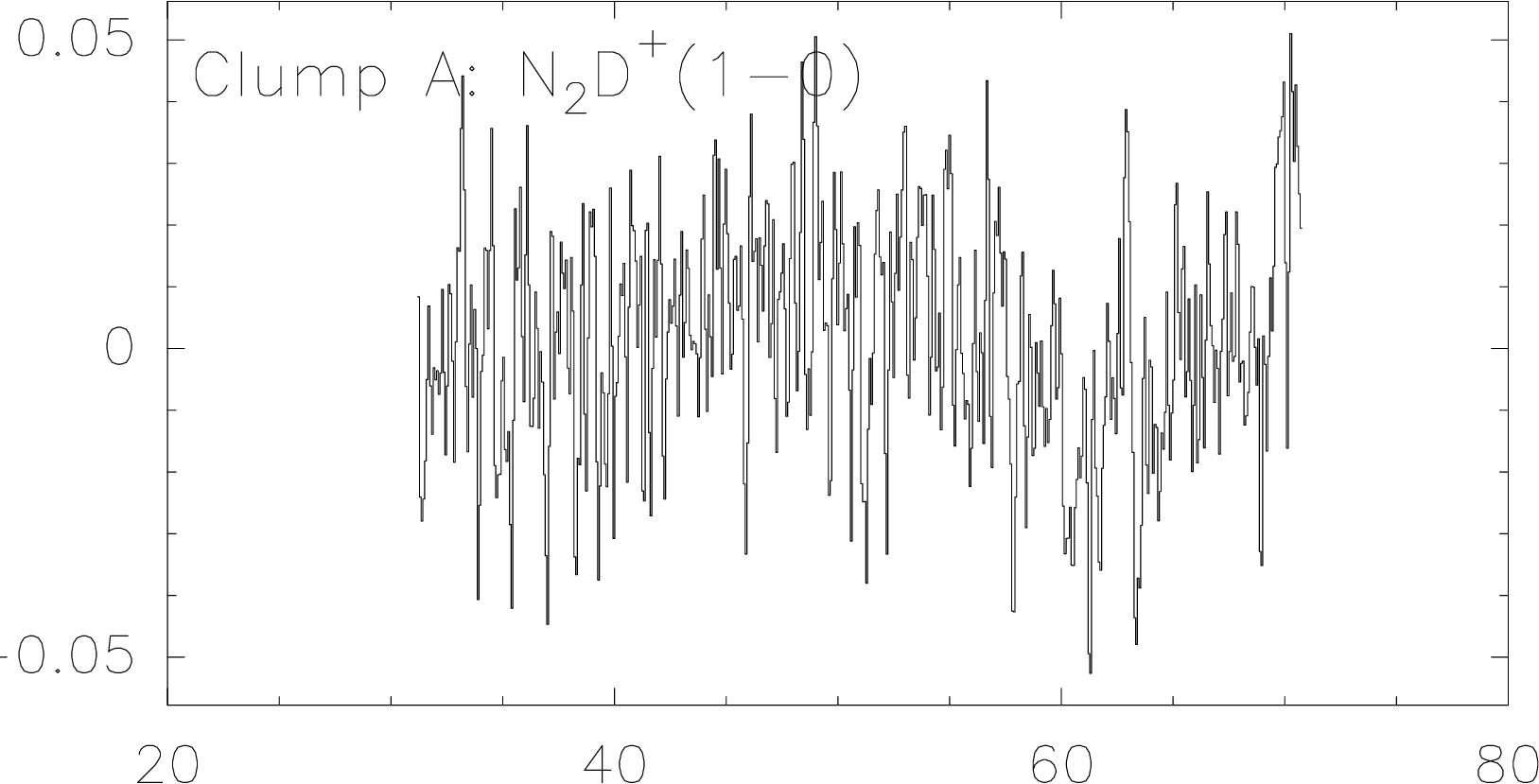}\\
\includegraphics[width=0.41\textwidth]{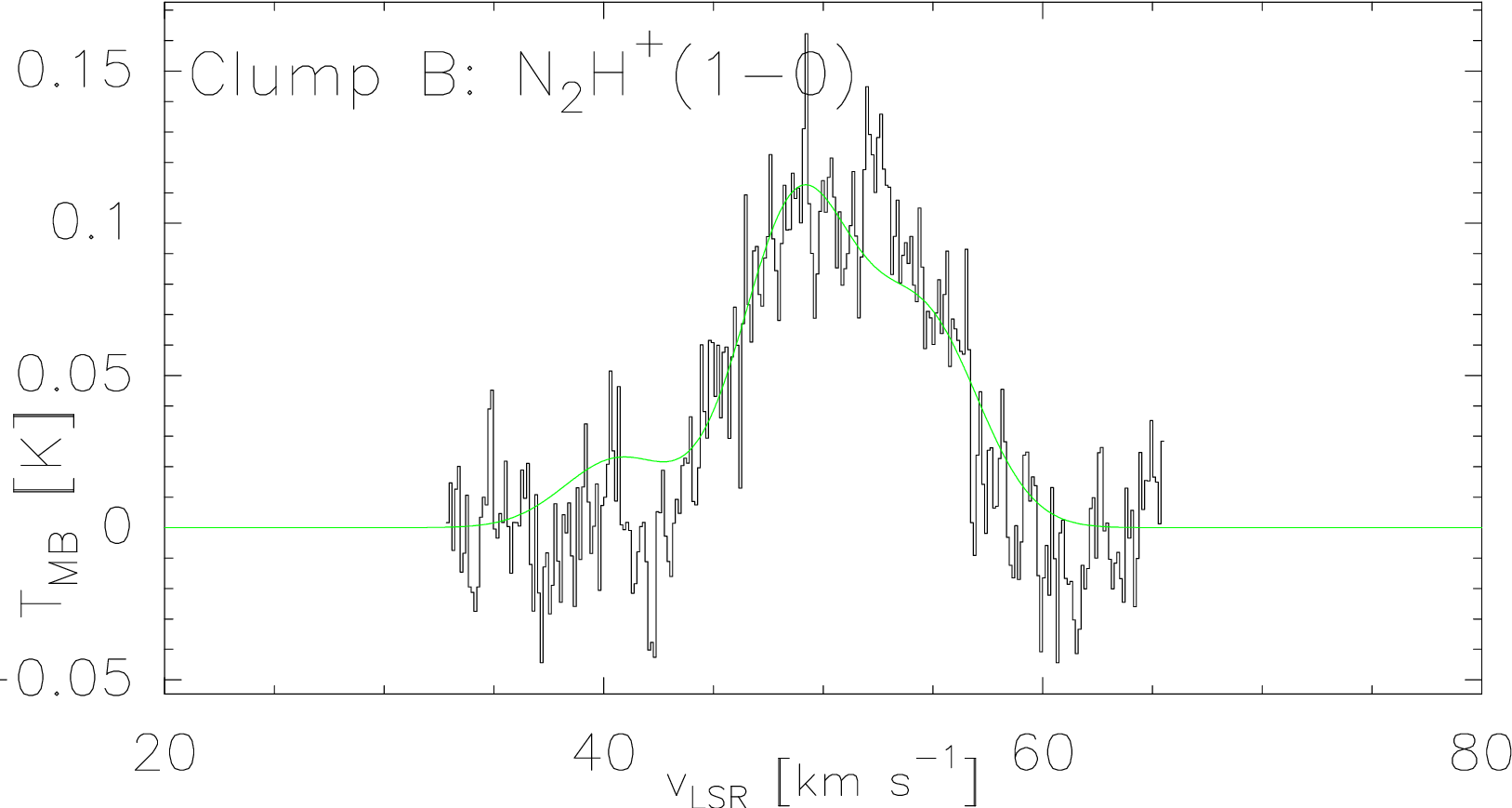}
\includegraphics[width=0.41\textwidth]{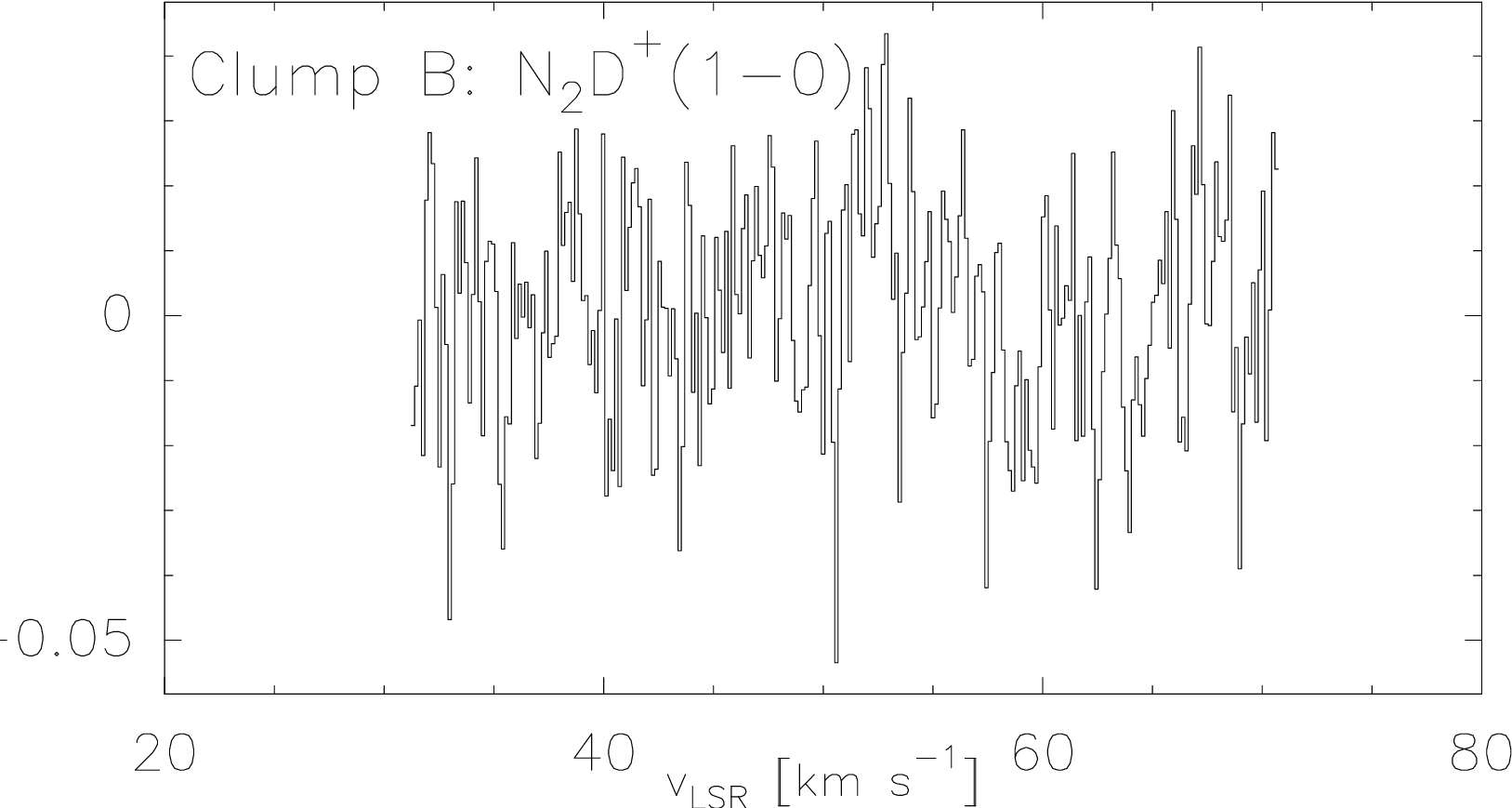}
\caption{The 
 spectra observed with the IRAM 30~m telescope. The~only detection is N$_2$H$^+(1-0)$ towards clump~B, where the hyperfine structure fit is shown in green. The~velocity range in the panels is different from that in Figures~\ref{figure:spectra1} and \ref{figure:spectra2} because of the narrower frequency bands used in the IRAM 30~m/VESPA~observations.\label{figure:spectra3}}
\end{figure}

To identify the spectral lines in the W band observed with the Yebes telescope,\linebreak we made use of the Splatalogue interface\endnote{\url{https://splatalogue.online/} accessed on 1 February 2024.
}, and~the Cologne Database for Molecular Spectroscopy (CDMS\endnote{\url{https://cdms.astro.uni-koeln.de/} accessed on 1 February 2024.
}; \cite{muller2005}) and the Jet Propulsion Laboratory (JPL) spectroscopic \linebreak database\endnote{\url{http://spec.jpl.nasa.gov/}
accessed on 1 February 2024. 
} \cite{pickett1998}. The~spectra of the spectral line transitions detected with Yebes towards both target clumps are shown in Figures~\ref{figure:spectra1} and \ref{figure:spectra2}. The~spectra observed with the IRAM 30~m telescope are shown in Figure~\ref{figure:spectra3}. The~detected spectral line transitions are listed in Table~\ref{table:lines}. As can be seen in Figures~\ref{figure:spectra1} and \ref{figure:spectra2}, the~Yebes spectra exhibit wavy baseline structures that could not be described by simple first- or second-order baseline functions. Wavy baselines can be the result of frequency switching that we employed in our Yebes observations (e.g.,~\cite{mangum2006, pagani2020}).

As shown in Figures~\ref{figure:spectra1} and \ref{figure:spectra2}, the~HCO$^+$ lines were fitted by a single Gaussian function using 
 {\tt CLASS90} (version jul21). The~HCN lines exhibit hyperfine splitting (three components in the case of the $J=1-0$ transition; e.g.,~\cite{loughnane2012}), and~we used the hyperfine structure method of {\tt CLASS90} to fit the detected lines. Owing to the detected HCN line profiles, the~fitting was conducted using a two-velocity component model. One could also interpret the observed HCN line profiles as asymmetric, where the redshifted peak is stronger than the blueshifted peak and where there is a self-absorption dip between the two peaks. This, however, would require that an optically thin line emission is seen at a velocity of about 27~km~s$^{-1}$ (estimated from the HCN spectrum towards clump~A), which is lower than the systemic local standard of rest (LSR) velocity of the cloud ($\sim$50~km~s$^{-1}$).

The HNCO lines also exhibit a hyperfine structure (e.g.,~\cite{velilla2015}), and~the observed $J=4-3$ transition has six hyperfine components. The~detected HNCO lines were fit using the hyperfine structure method and hyperfine component frequencies from the CDMS database. Again, the~fitting was made using two velocity components. Similar to the case of HCN, the~HNCO line detected towards clump~A could also be interpreted as a red asymmetric line profile with a central dip at about 36~km~s$^{-1}$.

We note that our original aim was also to observe the DCN$(J=1-0)$ transition at 72.415~GHz towards the target clumps (to study the [DCN]/[HCN] deuteration), but~the observed spectra had a very strong wave-like structure between about 72 and 76~GHz, which prevented any potential detection of the line.

The only line detected in our IRAM 30~m telescope observations is N$_2$H$^+(1-0)$ towards clump~B. Even this detection is relatively weak ($\sim$6$\sigma$) compared to the lines detected with Yebes. The~$J=1-0$ transition of N$_2$H$^+$ is split into 15 hyperfine components, which are mostly blended into one broad line in the observed spectra with a hint of a blended group at $\sim$40~km~s$^{-1}$. We fitted the hyperfine
structure in {\tt CLASS90} using the rest frequencies from~\cite{pagani2009} (Table~2 therein) and the relative intensities from
~\cite{mangum2015} (Table~8 therein).

The derived basic line parameters are listed in columns 2--5 in Table~\ref{table:parameters}. Besides~the formal $1\sigma$ fitting errors output by
{\tt CLASS90}, the~errors in the peak intensity ($T_{\rm MB}$) and the integrated intensity of the line ($\int T_{\rm MB} {\rm dv}$) also include the 10\% calibration uncertainty (the two sources of uncertainty were added in quadrature). We note that the quoted uncertainties in $\int T_{\rm MB} {\rm dv}$ should be interpreted as lower limits only because of the blended velocity components and ripples in the observed~spectra.

\unskip

\unskip

\unskip

\begin{table}[H] 
\caption{Detected spectral~lines.\label{table:lines}}
\newcolumntype{C}{>{\centering\arraybackslash}X}
\begin{tabularx}{\textwidth}{m{6cm}<{\centering}m{2cm}<{\centering}m{2cm}<{\centering}m{2cm}<{\centering}}
\toprule
\textbf{Transition} & \boldmath{$\nu$} & \boldmath{$E_{\rm u}/k_{\rm B}$} & \textbf{Telescope}\\
      & \textbf{[MHz]} & \textbf{[K]} \\
\midrule       
HNCO$(J_{K_a,\,K_c}=4_{0,\,4}-3_{0,\,3})$ & 87\,925.2178 \textsuperscript{a} & 10.55 & Yebes\\
HCN$(J=1-0)$ & 88\,631.846 \textsuperscript{b} & 4.25 & Yebes\\
HCO$^+(J=1-0)$ & 89\,188.5247 & 4.28 & Yebes\\
N$_2$H$^+(J=1-0)$ & 93\,173.7637 \textsuperscript{c} & 4.47 & IRAM 30~m\\
\bottomrule 
\end{tabularx}
{\footnotesize{The spectroscopic data were compiled from the CDMS database unless otherwise stated. In~columns 2 and 3, we list the line rest frequency and the upper-state energy divided by the Boltzmann constant. \textsuperscript{a} Frequency of the strongest hyperfine component $F=5-4$ (relative intensity $R_i = 0.4074$). \textsuperscript{b} Frequency of the strongest hyperfine component $F=2-1$ (relative intensity $R_i = 0.5555$) taken from Loughnane~et~al.~\cite{loughnane2012}. \textsuperscript{c} Frequency of the strongest hyperfine component $F_1,\,F = 2,\,3-1,\,2$ (relative intensity $R_i(F_1,\, J)R_i(F,\,F_1) = 7/27$; \cite{mangum2015}) taken from~\cite{pagani2009}.}}
\end{table}
\unskip

\subsection{Line Optical Thicknesses and Excitation~Temperatures}\label{sec3.2}

If a spectral line transition is split into hyperfine components, the~relative strengths of the
hyperfine components can be used to derive the line optical thickness, $\tau$. However, in~all cases where we fit the hyperfine structure of the line (Section~\ref{sec3.1}), the~hyperfine structure was not resolved (i.e., the separate components are blended), and hence, the corresponding optical
thicknesses should be taken with caution. When the optical thickness could be estimated by the {\tt CLASS90} routine, the~corresponding excitation temperature of the line was calculated using Equation~(1) in Miettinen~\cite{miettinen2020}. 

For HCO$^+$, we assumed that the rotational transition is thermalised at the dust temperature of the clump; that is, we assumed that $T_{\rm ex} = T_{\rm dust}$ (see column~4 in Table~\ref{table:G175sources}). The~values of $\tau$ and $T_{\rm ex}$ are listed in columns~6 and 7 in Table~\ref{table:parameters}.

\begin{table}[H]
\tablesize{\footnotesize}
\caption{Spectral 
 line parameters of the detected transitions and molecular column densities and fractional~abundances.\label{table:parameters}}
\newcolumntype{C}{>{\centering\arraybackslash}X}
\begin{adjustwidth}{-\extralength}{0cm}
\begin{tabularx}{\fulllength}{c c c c c c c c c c}
\toprule
\textbf{Transition} & \boldmath{${\rm v}_{\rm LSR}$} & \boldmath{$\Delta {\rm v}$} & \boldmath{$T_{\rm MB}$} & \boldmath{$\int T_{\rm MB} {\rm dv}$} & \boldmath{$\tau$} & \boldmath{$T_{\rm ex}$} & \boldmath{$N$} & \boldmath{$x$}\\
           & \textbf{[km~s\boldmath{$^{-1}$}]} & \textbf{[km~s\boldmath{$^{-1}$}]} & \textbf{[K]} & \textbf{[K~km~s\boldmath{$^{-1}$}]} & & \textbf{[K]} & \textbf{[cm\boldmath{$^{-2}$}]} & \\
\midrule
\underline{Clump A} & \\
HNCO$(4_{0,\,4}-3_{0,\,3})$ & $47.80\pm0.11$ & $8.28\pm0.26$ & $1.72\pm0.17$ & $35.09\pm3.53$ & $7.64\pm0.97$ & $4.7\pm0.2$ & $4.9\pm0.6(14)$ & $8.8\pm1.8(-9)$ \\
							& $32.10\pm0.28$ & $14.60\pm0.69$ & $0.99\pm0.10$ & $9.34\pm1.00$ & $0.12\pm0.01$ & $23.9\pm2.7$ & $1.9\pm0.2(14)$ & \ldots\\
HCN$(1-0)$ & $54.10\pm0.09$ & $10.90\pm0.35$ & $2.37\pm0.24$ & $35.07\pm3.55$ & $0.11\pm0.01$ & $43.2\pm5.4$ & $2.5\pm0.2(14)$ & $4.5\pm0.8(-9)$ \\
		   & $42.00\pm0.42$ & $43.90\pm0.83$ & $1.36\pm0.14$ & $64.31\pm6.40$ & $0.14\pm0.05$ & $21.4\pm6.5$ & $3.4\pm1.2(14)$ & \ldots\\
HCO$^+(1-0)$ & $53.10\pm0.16$ & $18.75\pm0.37$ & $1.47\pm0.15$ & $29.31\pm2.97$ & $0.14\pm0.01$\textsuperscript{a} & $14.3\pm0.4$ & $5.2\pm0.2(13)$ & $9.3\pm1.5(-10)$ \\

\underline{Clump B} & \\
HNCO$(4_{0,\,4}-3_{0,\,3})$ & $47.40\pm0.11$ & $14.70\pm0.29$ & $2.00\pm0.20$ & $33.36\pm4.96$ & $0.57\pm0.04$ & $12.8\pm1.1$ & $2.6\pm0.2(14)$ & $4.9\pm0.9(-9)$ \\
							& $35.50\pm0.04$ & $6.61\pm0.15$ & $1.10\pm0.11$ & $14.66\pm3.88$ & $11.20\pm0.81$ & $4.0\pm0.1$ & $5.2\pm0.4(14)$ & \ldots\\
HCN$(1-0)$ & $55.20\pm0.04$ & $11.90\pm0.28$ & $3.88\pm0.39$ & $79.85\pm8.05$ & $1.79\pm0.18$ & $9.3\pm0.7$ & $2.6\pm0.3(14)$ & $4.9\pm0.9(-9)$ \\
		   & $40.40\pm0.24$ & $36.70\pm0.23$ & $1.10\pm0.11$ & $68.06\pm6.90$ & $10.30\pm0.13$ & $4.0\pm0.1$ & $1.2\pm0.1(15)$ & \ldots\\
HCO$^+(1-0)$ & $53.80\pm0.07$ & $19.39\pm0.18$ & $2.69\pm0.27$ & $55.49\pm5.57$ & $0.27\pm0.02$\textsuperscript{a} & $14.5\pm0.5$ & $1.1\pm0.1(14)$ & $2.0\pm0.3(-9)$ \\
N$_2$H$^+(1-0)$ & $48.70\pm0.12$ & $5.79\pm0.21$ & $0.12\pm0.02$ & $1.14\pm0.11$ & $0.10\pm0.05$ & $7.8\pm2.5$ & $4.2\pm0.6(12)$ & $7.9\pm1.7(-11)$ \\
\bottomrule 
\end{tabularx}
\end{adjustwidth}
\noindent{\footnotesize{Columns 2–9 give the LSR radial velocity (${\rm v}_{\rm LSR}$), full width at half maximum (FWHM; 
$\Delta {\rm v}$), peak intensity ($T_{\rm MB}$), integrated intensity ($\int T_{\rm MB} {\rm dv}$ ), optical thickness ($\tau$), excitation temperature ($T_{\rm ex}$) of the line, and~the molecule's beam-averaged column density and fractional abundance with respect to H$_2$. The~latter two quantities are given in the form $a \pm b(c)$, which stands for $(a \pm b) \times 10^c$. \textsuperscript{a} The optical thickness was calculated under the assumption that $T_{\rm ex}=T_{\rm dust}$ (Section~\ref{sec3.2}).}}
\end{table}
\unskip

\subsection{Molecular Column Densities and Fractional~Abundances}

To calculate the beam-averaged column densities of the molecules, we used the standard local thermodynamic equilibrium (LTE) approach (see Equation~(32) in~\cite{mangum2015})\endnote{We note that for example in Miettinen~\cite{miettinen2020}, we used a column density formula, where the rotational degeneracy ($g_J$) does not appear in the denominator, while in Equation~(32) of Mangum \& Shirley~\cite{mangum2015} it does (see also their Equation~(33)). This difference arises from the different definitions of the dipole moment matrix element, which can be either $\lvert \mu_{\rm ul} \rvert^2=\mu^2 S$, where $\mu$ is the permanent electric dipole moment and $S$ is the line strength (Equation~(62) in~\cite{mangum2015}), or~$\lvert \mu_{\rm ul} \rvert^2=\mu^2 S/g_J$ (see~\cite{turner1991}, where $g_u=g_J$).}.

The values of the product $\mu^2 S$ were taken from the Splatalogue accessed CDMS database. The~rotational degeneracy ($g_J$) was calculated from the rotational quantum number of the upper state (see Equation~(34) in~\cite{mangum2015}). HNCO is an asymmetric top molecule and its $K$-level and reduced nuclear spin degeneracies ($g_K$ and $g_I$) were assigned values following the rules in Turner~\cite{turner1991} (see Appendix therein). The~value of $g_K$ equals 1 for asymmetric tops, and~the value of $g_I$ for HNCO is also 1 because the molecule has no identical interchangeable nuclei. The~partition functions were calculated using Equations~(3) and (4) in Miettinen~\cite{miettinen2014}.

The fractional abundances of the molecules were calculated by dividing the molecular column density by the H$_2$ column density. 
The H$_2$ column densities of the clumps were calculated from the LABOCA 870~$\upmu$m peak surfaces brightness and using the dust temperature given in column~4 in Table~\ref{table:G175sources} (see, e.g.,~Equation~(6) in~\cite{miettinen2020}). We assumed that the mean molecular weight per H$_2$ molecule is $\mu_{\rm H_2}=2.82$, the~dust opacity is $\kappa_{\rm 870\,\upmu m}=1.38$~cm$^2$~g$^{-1}$ at 870~$\upmu$m, and~the dust-to-gas mass ratio is $R_{\rm d/g}=1/141$ (see~\cite{miettinen2020} and references therein). The~angular resolution of our LABOCA data, $\sim$20 arcsec, is very similar to the Yebes 40~m telescope angular resolution at the frequency of the analysed spectral lines (i.e., \mbox{19.7--20~arcsec}), and~also comparable to the IRAM 30~m telescope beam size ($26.4$~arcsec). Hence, no smoothing of the LABOCA data was performed for calculating the H$_2$ column densities. The~beam-averaged column densities and fractional abundances of the molecules with respect to H$_2$ are listed in the last two columns in Table~\ref{table:parameters}. The~fractional abundance is only calculated for the main velocity component under the assumption that the observed submillimetre dust emission originates only in the target~cloud.

\subsection{Virial Analysis of the Cloud and Its~Clumps}

Miettinen~et~al.~\cite{miettinen2022} derived a line mass of $1\,011\pm146$~$M_{\odot}$~pc$^{-1}$ for the G1.75-0.08 filament. The~present spectral line data allow us to revise the virial or critical line mass of the cloud (see, e.g.,~Equation~(12) in~\cite{fiege2000}). To~calculate the latter quantity, we used the total (thermal+non-thermal) velocity dispersion, where the observed spectral line width (FWHM) was taken to be the average of the FWHMs of the HNCO lines detected towards clump~A and clump~B, because~(i) the transition has a high critical density ($10^6$~cm$^{-3}$ at 20~K;~\cite{sanhueza2012}, Table~1 therein), (ii) the detected lines do not exhibit as strong wing emission as the HCN lines, and~(iii) HNCO was detected towards both clumps. The~value of the aforementioned average FWHM is $11.5\pm3.2$~km~s$^{-1}$. We assumed that the gas kinetic temperature is equal to the dust temperature ($15.0\pm0.4$~K for the filament;~\cite{miettinen2022}). The~derived virial line mass and the ratio of the observed and virial line masses are listed in Table~\ref{table:virial}.

We also revised the virial masses and virial parameters (the ratio of the virial mass to the source mass) of the target clumps derived by Miettinen~et~al.~\cite{miettinen2022} (see Section~3.5 therein). For~this purpose, we also employed the HNCO line widths. It was again assumed that the gas temperature is equal to the dust temperature. The~clumps were assumed to have a radial density profile of the form $n(r)\propto r^{-1.6}$, which is consistent with those derived for Galactic high-mass star-forming clumps (see~\cite{miettinen2020b} and references therein). The power-law index of the density profile, $p$, modifies the virial mass as $M_{\rm vir}\propto a^{-1}$, where $a=(1-p/3)/(1-2p/5)$ (see~\cite{miettinen2012b} and references therein). The~mean molecular weight per free particle was assumed to be $\mu_{\rm p}=2.37$. The~derived virial masses and virial parameters of the clumps are listed in Table~\ref{table:virial}.

\begin{table}[H] 
\caption{Virial masses and virial parameters of the target~sources.\label{table:virial}}
\newcolumntype{C}{>{\centering\arraybackslash}X}
\begin{tabularx}{\textwidth}{CCC}
\toprule
\textbf{Source} & \boldmath{$M_{\rm vir}$} & \boldmath{$\alpha_{\rm vir}$} \textbf{\textsuperscript{a}} \\
\midrule
G1.75-0.08 & $11,113\pm 6171$ M$_{\odot}$~pc$^{-1}$ \textsuperscript{b} & $0.09\pm0.05$ \\
Clump A & $16,248\pm1017$ M$_{\odot}$ & $3.0\pm0.5$\\
Clump B & $43,384\pm1710$ M$_{\odot}$ & $11.5\pm2.3$ \\
\bottomrule 
\end{tabularx}
\newline
\noindent{\footnotesize{\textsuperscript{a} For the G1.75-0.08 filament, the~virial parameter is defined as the ratio of the line mass to virial line mass, while for the clumps the virial parameter is defined the other way round, that is, as the ratio of the virial mass to the clump mass.} \textsuperscript{b} Virial or critical line mass.}
\end{table}
\unskip

\section{Discussion}\label{sec4}
\unskip

\subsection{Spectral line Profiles and Cloud~Kinematics}

The HCN and HNCO lines we detected towards the clumps of G1.75-0.08 show asymmetric profiles that we have interpreted to be caused by two different velocity components {(see Figures~\ref{figure:spectra1} and \ref{figure:spectra2})}. The~presence of multiple nearby velocity components complicates the interpretation of the gas kinematics of the cloud. In~principle, the~detected red asymmetric line profiles could also be indicative of outward motions or cloud expansion (e.g.,~\cite{gregersen1997, gao2010, kristensen2012, qin2016}). For~this to be the case, however, we should see optically thin line emission at a systemic velocity that matches the velocity of the central dip, which is not the case. Hence, the~presence of multiple different velocity components seems more likely. Indeed, physically unassociated clouds along the line of sight that have different LSR velocities would not be unexpected owing to the large distace of G1.75-0.08. However, this needs to be tested by further spectral line~observations.

The HCN$(1-0)$ spectrum extracted towards the LABOCA peak position of clump~A by Miettinen~\cite{miettinen2014} also showed a hint of a red asymmetric profile (see Figure~C.6 therein).\endnote{Clump~A and clump~B were called SMM~15 and SMM~8 in the target field G1.87-014 of Miettinen~\cite{miettinen2014} (see Figure~1 therein).} The HNCO$(4_{0,\,4}-3_{0,\,3})$ line detected by Miettinen~\cite{miettinen2014} towards clump~A was interpreted as a single, very broad line ($30.40 \pm 1.39$~km~s$^{-1}$ in FWHM), but~our 1.9 times higher angular resolution and more sensitive observations have revealed the presence of two velocity components. The~HCO$^+(1-0)$ spectrum towards clump~A in Miettinen~\cite{miettinen2014} was interpreted to exhibit two velocity components, while in the present study the line appears to have only one clear velocity component. The~larger beam of the MALT90 observations (38~arcsec) might have captured emission from another velocity component that is now avoided. However, the~potential presence of an additinal component at $\sim$30~km~s$^{-1}$ is blurred by the rippled baseline in the HCO$^+(1-0)$ spectrum.

All the MALT90 spectra towards clump~B in Miettinen~\cite{miettinen2014} were detected to have only one velocity component (see Figure~C.2 therein). The~difference compared to the present results is most notable for HCN, where we now see a broad ($36.70 \pm 0.23$~km~s$^{-1}$ in FWHM) secondary component. The~HNCO line we detected with Yebes, however, could have been interpreted to have a single-velocity component as conducted by Miettinen~\cite{miettinen2014}, but~in that case, the radial velocity of the line would have been lower (44.5~km~s$^{-1}$) than that derived for the stronger peak in the HCN spectrum (55.2~km~s$^{-1}$), which presumably originates in our target cloud. The~N$_2$H$^+(1-0)$ line towards clump~B analysed in Miettinen~\cite{miettinen2014} was derived from having an FWHM of $20.10\pm1.69$~km~s$^{-1}$, which is $3.5\pm0.3$ broader than derived in the present study. In~this case, the~larger beam of the MALT90 observations might have captured emission from gas with higher velocity dispersion than probed by our new IRAM~observations.

Miettinen~et~al.~\cite{miettinen2022} speculated that G1.75-0.08 might represent a case where a filamentary cloud is undergoing gravitational focussing or the so-called edge effect, where gas clumps have accumulated at both ends of the filament (e.g.,~\cite{burkert2004, heigl2022, hoemann2023a, hoemann2023b}). It would be tempting to interpret the detected HCN and HNCO line profiles as red asymmetric lines that indicate the presence of outward gas motions, because~this might support the hypothesis of gravitational focussing. However, further spectral line observations are needed to test this~hypothesis.

CMZ clouds can have a complex velocity field, as was demonstrated by Henshaw~et~al.~\cite{henshaw2019} in the case of the CMZ cloud G0.253+0.016 (also known as the Brick). Henshaw~et~al.~\cite{henshaw2019} found that the Brick is not a single, coherent cloud, but~rather a structured system of different velocity components and complex dynamics that might be the result of the orbital dynamics and shear motions in the CMZ. On~the basis of the present results, G1.75-0.08 could be an analogue object with the Brick. High angular and spectral resolution imaging of G1.75-0.08 would, however, be required to reach a better understanding of the velocity structure of the~cloud.

\subsection{Dynamical State of~G1.75-0.08}\label{sec4.2}

Using the HCN$(1-0)$ line width (FWHM) of $13.50\pm0.38$~km~s$^{-1}$ detected in the MALT90 survey, Miettinen~et~al.~\cite{miettinen2022} derived a low value of $0.07\pm0.01$ for the ratio of the line mass to the critical line mass for G1.75-0.08. In~the present paper, we have used our new molecular line data (specifically HNCO line widths) to revise the latter value to $0.09\pm0.05$ (Table~\ref{table:virial}), which is very close to the earlier estimate. Hence, our finding supports the view that G1.75-0.08 is strongly subcritical (by a factor of $11\pm6$), which makes it very different compared to the general population of Galactic filaments for which the line mass and the critical line mass are often found to agree within a factor of $\sim$2 (e.g.,~\cite{mattern2018}; Figure~27 therein). One could speculate that G1.75-0.08 is subject to tidal disruption effects near the Galactic Centre ($R_{\rm GC}$$\sim$270~pc;~\cite{miettinen2022}). The~dynamical CMZ environment is found to have an influence on the Brick~\cite{henshaw2019}, and~hence, this might be the case for G1.75-0.08 as~well.

We note that if the FWHM of N$_2$H$^+(1-0)$ detected towards clump~B would be used in the analysis, the~virial parameter for G1.75-0.08 would become $\alpha_{\rm vir}^{\rm fil}=0.36\pm0.06$, in~which case the filament would be subcritical only by a factor of $2.8\pm0.5$. However, N$_2$H$^+(1-0)$ is detected only in clump~B, and the detection is relatively weak compared to other line detections;~hence, the corresponding line FWHM might not be as reliable as for HNCO used in the~analysis.

\subsection{Dynamical State of the~Clumps}

Our new molecular line data also allowed us to revise the virial parameters of the clumps in G1.75-0.08. Miettinen~et~al.~\cite{miettinen2022} found that clumps~A and B are both gravitationally unbound with $\alpha_{\rm vir} \gg 2$ (see Table~7 therein). Our new data support the conclusion that the clumps are gravitationally unbound ($\alpha_{\rm vir} > 2$; Table~\ref{table:virial}), although~clump~A lies only a factor of $1.5\pm0.3$ away from being gravitationally~bound. 

We note that if the FWHM of N$_2$H$^+(1-0)$ detected towards clump~B would be used in the calculation, the~virial parameter of clump~B 
would become $\alpha_{\rm vir}=1.8\pm0.4$, which would suggest that the clump is marginally gravitationally bound. However, as~noted in Section~\ref{sec4.2}, 
the N$_2$H$^+(1-0)$ detection towards clump~B is relatively weak and the corresponding line FWHM should be interpreted with caution. It should also be noted that further support against gravity is provided by a magnetic field, and~the magnetic field strength in the CMZ is known to be relatively strong compared to molecular clouds at larger Galactocentric distances (e.g.,~\cite{tress2024}).

The clumps appear dark at 70~$\upmu$m and lie above the mass--radius threshold for high-mass star formation proposed by Baldeschi~et~al.~\cite{baldeschi2017}, as shown in Figure~7 of a study by Miettinen~et~al.~\cite{miettinen2022}, and which is given by $M_{\rm thresh}=1\,732~{\rm M_{\odot}}\times(R/{\rm pc})^{1.42}$ when scaled to our assumptions about the dust opacity and gas-to-dust mass ratio~\cite{miettinen2020b}. However, the~present data do not suggest the clumps to be candidates for being high-mass prestellar clumps, but~only high-mass starless clumps. Moreover, 70~$\upmu$m darkness does not necessarily mean that the clump is quiescent or devoid of star formation, and~such objects can host embedded low- and intermediate-mass protostellar cores (e.g.,~\cite{traficante2017, sanhueza2019, li2020}). For~example, the~line wing emission seen in our HCN and HNCO spectra might arise from protostellar outflow activity. However, this is only speculation and requires high-resolution spectral line imaging to be confirmed or disproved. On the other hand, a~deficit of gravitationally bound clumps in the CMZ could explain its low star formation rate (SFR) compared to the Galaxy in general~\cite{myers2022}. Chabrier and Dumond~\cite{chabrier2024} suggested that molecular clouds in the CMZ are subject to only one episode of large-scale turbulence injection during their lifetime, where the injection is mostly provided by the gas inflow driven by the Galactic bar. Hence, there can be less injection of turbulence and turbulent motion will eventually decay, which facilitates the formation of stars. This could then explain the low SFR compared to the Galactic disc. A low SFR in the CMZ may also be related to the tidal effects of the Galactic Centre.

Dust continuum imaging of G1.75-0.08 with ArT\'eMiS by Miettinen~et~al.~\cite{miettinen2022} revealed that clumps~A and B show substructure (see Figure~8 therein). Dense cores in gravitationally unbound clumps have also been observed in other IRDCs (e.g., G340.222–00.167 with $\alpha_{\rm vir}=5.7\pm1.7$; \cite{li2023}). If~the substructure detected with ArT\'eMiS is physical, it could be an indication that the inner parts of the clump have decoupled from the more turbulent outer parts of the clump and that gravitational fragmentation has taken place in the denser region within the parent~clump.

The observed, projected separation of the substructures in the clumps is a factor of $1.5\pm0.1$ larger than the thermal Jeans length for clump~A and a factor of $1.3\pm0.1$ larger in clump~B (see Table~8 in~\cite{miettinen2022}). Our new spectral line data (HNCO) suggest that the Jeans lengths that take the non-thermal motions into account would be larger by factors of about 10.5 (clump~A) and 21 (clump~B) than the observed substructure separation. Hence, the~observed substructure in the clumps is roughly consistent with thermal Jeans~fragmentation.

\subsection{Molecular Detections and Abundances in~G1.75-0.08}

\subsubsection{HNCO (Isocyanic Acid)}

Based on the assumption that the higher velocity component of the HNCO line is associated with G1.75-0.08, we derived an HNCO fractional abundance of $(8.8\pm1.8)\times 10^{-9}$ towards clump~A and $(4.9\pm0.9)\times 10^{-9}$ towards clump~B. 

For comparison, Vasyunina~et~al.~\cite{vasyunina2011}, who used the 22~m Mopra telescope, detected the HNCO$(4_{0,\,4}-3_{0,\,3})$ transition in 13 out of their sample of 37 clumps in 15 different IRDCs (35\% detection rate), and~derived abundances in the range of $(0.17-2.86)\times 10^{-9}$. The~latter values were scaled down by a factor of 0.7735 to take the different assumptions used in the calculation into account (e.g., the dust opacity and dust-to-gas mass ratio), which is needed for a proper comparison with our results (see~\cite{miettinen2020} for details). The~abundances we derived towards our target clumps are $3.1\pm0.6$ (clump~A) and $1.7\pm0.3$ (clump~B) times higher than the highest value in the Vasyunina~et~al.~\cite{vasyunina2011} sample. We note that the distances of the Vasyunina~et~al.~\cite{vasyunina2011} target sources lie in the range of 2.1--5.3~kpc, and~hence none of their sources are associated with the CMZ. Nevertheless, the~HNCO abundances are not significantly different from those in~G1.75-0.08.

Sanhueza~et~al.~\cite{sanhueza2012}, who also used the 22~m Mopra telescope, detected HNCO$(4_{0,\,4}-3_{0,\,3})$ in 18 of their sample of 92 clumps in IRDCs (20\% detection rate), and~derived abundances in the range of $(0.15-4.45)\times 10^{-9}$. The~latter values were scaled down by
a factor of 0.549 for a more meaningful comparison with our results (see~\cite{miettinen2020}). The~highest HNCO abundance in the Sanhueza~et~al.~\cite{sanhueza2012} sample is similar to that in clump~B (agreement within a factor of $1.1\pm0.2$). 

We note that the HNCO$(4_{0,\,4}-3_{0,\,3})$ line has also been detected in the Brick as part of the MALT90 survey~\cite{rathborne2014} and also with the Atacama Large Millimetre/submillimetre Array (ALMA) \cite{rathborne2015, henshaw2019}, but~no fractional abundance estimate was presented to compare~with.

\subsubsection{HCN (Hydrogen Cyanide)}

The HCN fractional abundances we derived for clump~A and clump~B are comparable to each other ($(4.5\pm0.8)\times 10^{-9}$ and $(4.9\pm0.9)\times 10^{-9}$, respectively). 

Vasyunina~et~al.~\cite{vasyunina2011} detected HCN$(1-0)$ in all of their 37 clumps, and~the scaled fractional abundances lie in the range of $(0.26-5.26)\times 10^{-9}$. The~latter range brackets the HCN abundances we derived. Sanhueza~et~al.~\cite{sanhueza2012} also reported a high detection rate of HCN$(1-0)$ for their sample (80\%), but~the fractional abundances were not derived owing to blended hyperfine components and complex line~profiles.

\subsubsection{HCO$^+$ (Formyl Ion)}

The HCO$^+$ abundance towards clump~A was derived to be $(9.3\pm1.5)\times 10^{-10}$, while a value of $(2.0\pm0.3)\times 10^{-9}$ was derived for clump~B.

Vasyunina~et~al.~\cite{vasyunina2011} detected HCO$^+(1-0)$ in 31 clumps (83.8\% detection rate), and~the scaled fractional abundances lie in the range of $(0.27-3.94)\times 10^{-8}$. The~lowest value in this range is $2.9\pm0.5$ and $1.4\pm0.2$ times higher than the abundance we derived for clump~A and clump~B. Sanhueza~et~al.~\cite{sanhueza2012} reported a comparably high detection rate of HCO$^+(1-0)$ for their sample (88\%), and~the scaled values of the fractional abundances lie in the range of $(0.21-15.3)\times 10^{-8}$. Again, the~HCO$^+$ abundance in our clumps is closer to the lower end of values in the Sanhueza~et~al.~\cite{sanhueza2012} sample. The~authors found that the HCO$^+$ abundance increases as the clump evolves, which in turn could be related to the increasing temperature, which leads to the release of CO from the icy grain mantles into the gas phase out of which HCO$^+$ can then primarily form in the reaction with H$_3^+$. Because~our target clumps are 70~$\upmu$m dark and apparently quiescent, their low HCO$^+$ abundances are consistent with the aforementioned evolutionary~trend.

\subsubsection{N$_2$H$^+$ (Diazenylium)}

We detected N$_2$H$^+$ only in clump~B and derived an abundance of $(7.9\pm1.7)\times 10^{-11}$ for the species. For~comparison, Vasyunina~et~al.~\cite{vasyunina2011} detected N$_2$H$^+(1-0)$ in all except one of their clumps (97.3\% detection rate), and~the scaled fractional abundances lie in the range of $(0.15-7.74)\times 10^{-9}$. Our derived N$_2$H$^+$ abundance is roughly comparable to the lowest value in this range (a factor of $1.9\pm0.4$ difference). Sanhueza~et~al.~\cite{sanhueza2012} reported a very high detection rate of 97\% for the strongest hyperfine component of N$_2$H$^+(1-0)$ (Table~4 therein), and~the fractional abundances they derived are $(0.1-9.2)\times 10^{-9}$. The~lowest value in this range agrees with the abundance of N$_2$H$^+$ in clump~B within a factor of $1.3\pm0.3$. The~authors found that the N$_2$H$^+$ abundance increases with clump evolution, and~the low N$_2$H$^+$ abundance found for the quiescent clump~B is consistent with this evolutionary trend. The~physics and chemistry behind this trend might be related to the release of N$_2$ from dust grains as the temperature in the clump increases, after~which it can react with H$_3^+$ to form N$_2$H$^+$. This process is competing with the increasing abundance of CO, which is destroying N$_2$H$^+$ in a process that produces HCO$^+$ and N$_2$.

\subsubsection{[N$_2$H$^+$]/[HCO$^+$] Abundance Ratio and [N$_2$D$^+$]/[N$_2$H$^+$] Deuteration}

Sanhueza~et~al.~\cite{sanhueza2012} found that the [N$_2$H$^+$]/[HCO$^+$] ratio can be used as a chemical clock, where the ratio decreases as the clump evolves from the intermediate stage (with enhanced 4.5~$\upmu$m emission or an embedded 24~$\upmu$m source) via an active stage (both enhanced 4.5~$\upmu$m emission and a 24~$\upmu$m source) to the red clump stage (association with bright 8~$\upmu$m emission). However, quiescent or IR-dark clumps were not found to follow this trend (see Figure~16 in their study). The~[N$_2$H$^+$]/[HCO$^+$] abundance ratio for clump~B is $0.04\pm0.01$ (calculated from the column density ratio), which is two times lower than the median value of 0.08 derived by Sanhueza~et~al.~\cite{sanhueza2012} for their quiescent clumps, and~also lower than the median value of 0.07 the authors derived for red clumps. Hence, our quiescent clump~B also does not appear to follow the trend seen in the median values of [N$_2$H$^+$]/[HCO$^+$] by Sanhueza~et~al.~\cite{sanhueza2012}.

Using a $3\sigma$ intensity upper limit of $T_{\rm MB}<4$~mK for N$_2$D$^+(1-0)$ towards clump~B, and~assuming the same line width (FWHM) and $T_{\rm ex}$ as for the detected N$_2$H$^+(1-0)$ line, we derived an N$_2$D$^+$ column density upper limit of $<$1.8 $\times$ $10^{11}$~cm$^{-2}$. This suggests that the [N$_2$D$^+$]/[N$_2$H$^+$] deuterium fractionation in clump~B is $<0.05$. For~comparison, using observations with the IRAM 30~m telescope, Fontani~et~al.~\cite{fontani2006} derived [N$_2$D$^+$]/[N$_2$H$^+$] values in the range of $\leq$0.004$-$0.02 for their sample of ten high-mass young stellar objects. Fontani~et~al.~\cite{fontani2011} studied a sample of high-mass starless cores, high-mass protostellar objects (HMPOs), and~ultracompact H{\scriptsize II} regions, and~found [N$_2$D$^+$]/[N$_2$H$^+$] ratios of \mbox{0.012--0.7}, 0.017-- $\leq$ 0.4, and~0.017-- $\leq$ 0.08 for these different evolutionary stages. The~average deuteration values were found to be $\sim$0.26, 0.037, and~0.044, respectively, showing a decreasing trend in [N$_2$D$^+$]/[N$_2$H$^+$] when the source evolves from a starless stage to a HMPO. Gerner~et~al.~\cite{gerner2015} found that the [N$_2$D$^+$]/[N$_2$H$^+$] ratio in IRDCs drops from a median value of about 0.032 to about 0.009 in HMPOs (see Figure~5 therein). Based on observations with APEX, Lackington~et~al.~\cite{lackington2016} derived the [N$_2$D$^+$]/[N$_2$H$^+$] ratios between 0.002 and 0.23 for their sample of 29 cores in IRDCs. The~[N$_2$D$^+$]/[N$_2$H$^+$] upper limit we derived for the 70~$\upmu$m dark clump~B is consistent with many of the deuteration levels observed in massive clumps and other IRDCs, but~the present data do not allow us to quantify the potential effect that the environment might have on the deuteration in G1.75-0.08 (e.g., is it exceptionally low compared to the IRDCs in other parts of the Galaxy).

\section{Conclusions}\label{sec5}

We used the Yebes 40~m and IRAM 30~m telescopes to make the first single-pointed spectral line observations towards the IRDC G1.75-0.08. These new observations were used to study the kinematics, dynamics, and~molecular abundances of the cloud and its clumps. These~new data allowed us to revise the gas velocity dispersion-dependent physical properties of the target source. Our main results are summarised as follows:

   \begin{enumerate}
      \item Three different molecular line transitions were unambiguously detected towards the clumps in G1.75-0.08 with Yebes, namely, HNCO$(J_{K_a,\,K_c}=4_{0,\,4}-3_{0,\,3})$, HCN$(J=1-0)$, and~HCO$^+(J=1-0)$. With~the IRAM 30~m telescope, we detected only N$_2$H$^+(J=1-0)$ towards clump~B. 
	  \item The HCN and HNCO spectra exhibit two velocity components, which give an impression of red asymmetric line profiles that would be an indication of expanding gas motions.
      \item Our new spectral line data support the view that the G1.75-0.08 filament is strongly subcritical (by a factor of $11\pm6$), which is atypical compared to the general population of Galactic molecular cloud filaments.
	  \item Both clumps at the ends of the G1.75-0.08 filament were found to be gravitationally unbound ($\alpha_{\rm vir}>2$). Because~the clumps are 70~$\upmu$m dark and massive (several $10^3$~M$_{\odot}$), they can be considered candidates for being high-mass starless clumps, but~not prestellar.
	  \item The fractional abundances of the detected species in the target clumps are consistent with those observed in other IRDCs.
   \end{enumerate}

The IRDC G1.75-0.08 lies about 270~pc from the Galactic Centre in the CMZ, and~could be an analogue of the CMZ cloud G0.253+0.016 (the Brick), which has been found to be a dynamically complex and hierarchically structured system rather than a single, coherent cloud~\cite{henshaw2019}. High-resolution spectral line imaging of G1.75-0.08 would be needed to quantify the cloud's velocity structure; examine whether the orbital dynamics and shear motions in the CMZ could affect the cloud, as suggested in the case of the Brick; and~test the hypothesis that the origin of the two clumps at the ends of the filament could be the result of gravitational focussing or the edge~effect.



\vspace{6pt} 




\authorcontributions{Conceptualisation, O.M.; observations, O.M. and M.S.-G.; data reduction, O.M. and M.S.-G.; methodology, O.M. and M.S.-G.; formal analysis, O.M.; investigation, O.M. and M.S.-G.; data curation, O.M. and M.S.-G.; writing---original draft preparation, O.M.; writing---review and editing, O.M. and M.S.-G.; visualisation, O.M. All authors have read and agreed to the published version of the manuscript.}

\funding{This research received no external~funding.}

\institutionalreview{Not applicable.}

\informedconsent{Not applicable.}

\dataavailability{The Yebes and IRAM spectral line data that support the findings of this study are available upon request from the corresponding author. The~APEX dust continuum data are openly accessible online through the ESO Science Archive (\url{https://archive.eso.org/wdb/wdb/eso/apex/form}
).} 




\acknowledgments{We thank the two anonymous reviewers for providing useful comments and suggestions that helped to improve the quality
of this paper. We are grateful to the staff at the Yebes 40~m telescope for performing the service mode observations for project 23A004 and~to the staff at the IRAM 30 m telescope for performing the service mode observations presented in this paper. This research has made use of NASA's Astrophysics Data System Bibliographic Services, the~NASA/IPAC Infrared Science Archive, which is operated by the Jet Propulsion Laboratory, California Institute of Technology, under~contract with the National Aeronautics and Space Administration, and~{\tt~Astropy}\endnote{\url{www.astropy.org} accessed on 1 April 2023. 
}, a~community-developed core Python package for Astronomy~\cite{astropy2013, astropy2018}.}

\conflictsofinterest{The authors declare no conflicts of~interest.} 



\abbreviations{Abbreviations}{
The following abbreviations are used in this manuscript:\\

\noindent 
\begin{tabular}{@{}ll}

ALMA & Atacama Large Millimetre/submillimetre Array \\
APEX & Atacama Pathfinder EXperiment \\
ArT\'eMiS & Architectures de bolom\`{e}tres pour des T\'elescopes \`{a} grand champ \\
          & de vue dans le domaine sub-Millim\'etrique au Sol \\
CDMS & Cologne Database for Molecular Spectroscopy \\
CLASS & Continuum and Line Analysis Single-dish Software \\
CMZ & Central Molecular Zone \\
EMIR & Eight MIxer Receiver \\
FFTS & Fast-Fourier Transform Spectrometer\\
FWHM & Full width at half maximum \\
GILDAS & Grenoble Image and Line Data Analysis Software \\
HMPO & High-mass protostellar object \\
HPBW & Half-power beam width \\
IPAC & Infrared Processing \& Analysis Center \\
IRAM & Institut de Radioastronomie Millim\'etrique \\
IRDC & Infrared dark cloud \\
JPL & Jet Propulsion Laboratory \\
LABOCA & Large APEX BOlometer CAmera \\
LSR & Local standard of rest \\
LTE & Local thermodynamic equilibrium \\
MALT90 & Millimetre Astronomy Legacy Team 90 GHz \\
NASA & National Aeronautics and Space Administration \\
PWV & Precipitable water vapour \\
SFR & Star formation rate \\
VESPA & Versatile SPectrometer Array \\
\end{tabular}
}




\begin{adjustwidth}{-\extralength}{0cm}
\printendnotes[custom] 

\reftitle{References}

\PublishersNote{}
\end{adjustwidth}
\end{document}